\documentclass[preprint]{aastex} 
\def\arcmin{$^\prime$}
\def\arcsec{$^{\prime\prime}$}
\newcommand{\HI}{\protect\ion{H}{1}}
\newcommand{\msun}{$M_\odot$}

\newcommand{\mhi}{$M_{HI}$}

\newcommand{\kms}{~km\,s$^{-1}$}

\shorttitle{Anomalous \HI\ in NGC 2997} 
\shortauthors{Hess et al.}
\begin{document}

\title{Anomalous \HI\ in NGC~2997}

\author{Kelley M. Hess\altaffilmark{1}} \affil{Department of Astronomy,
  University of Wisconsin-Madison, Madison, WI 53706;\\
  hess@astro.wisc.edu}

\author{D.~J. Pisano\altaffilmark{2}}
\affil{WVU/NRAO, WVU Department of Physics, P.O. Box 6515, Morgantown, WV 26506-6315;\\
  dpisano@nrao.edu}

\author{Eric M. Wilcots\altaffilmark{1}} \affil{Department of Astronomy,
  University of Wisconsin-Madison, Madison, WI 53706;\\
  ewilcots@astro.wisc.edu}

\and

\author{Jayaram N. Chengular\altaffilmark{3}} \affil{NCRA (TIFR) Pune 411007
  India;\\ chengalur@ncra.tifr.res.in}

\begin{abstract}
  We present deep \HI\ observations of the moderately inclined spiral galaxy,
  NGC~2997.  The goal of these observations was to search for \HI\ clouds in
  the vicinity of NGC 2997 analogous to the high velocity clouds of the Milky
  Way and gain insight into their origins.  We find evidence for the presence
  of a galactic fountain as well as the accretion of intragalactic material,
  however we do not identify any large clouds of \HI\ far from the disk of the
  galaxy.  NGC~2997 has a thick, lagging \HI\ disk that is modeled with a
  vertical velocity gradient of $18-31$\kms\ kpc$^{-1}$.  Anomalous velocity \HI\
  clouds with masses of order $10^7$ \msun, which cannot be explained by galactic
  fountain models allow us to estimate a lower limit to the accretion of
  extragalactic gas of 1.2 \msun\ yr$^{-1}$.  The number and mass of these clouds
  have implications for cosmological simulations of large scale structure and
  the presence of dark matter halos.  We have used values from the
  the literature to estimate a star formation rate of $5\pm1$ \msun\ yr$^{-1}$ and 
  to derive a new distance to NGC~2997 of $12.2\pm0.9$ Mpc using published
  Tully-Fisher relations.
\end{abstract}


\section{Introduction}
In the Milky Way Galaxy, 21 cm observations of neutral atomic hydrogen (\HI) provide
a unique probe of the structure and dynamics of the interstellar medium.  
\citet{1963ASCP} and others have identified clumps of \HI\ emission with 
motions that are inconsistent with Galactic rotation.  Formally, these high 
velocity clouds (HVCs) are identified by either having a line of sight velocity, 
$|v_{LSR}|$, greater than 90\kms\ with respect to the local standard of rest or else
differ from the rotational velocity of the disk by greater than 50\kms
at a given Galactic longitude \citep{1997ARA&A..35..217W}.  Despite over 40
years of study, the origin of these HVCs remains a mystery.  Our position 
within the Milky Way's disk provides a unique perspective of their spatial 
distribution.  However, because the clouds do not obey Galactic rotation, distance 
measurements rely on the serendipitous position of stars and stellar 
spectroscopy to find distance brackets \citep[e.g.][]{2006ApJ...638L..97T,
2007arXiv0712.0612T,2007ApJ...670L.113W,2008ApJ...672..298W} or other 
coincidences, such as the association of the Magellanic Stream with
the Magellanic Clouds, or the interaction of Smith's Cloud 
\citep{2008ApJ...679L..21L} and the Magellanic Stream \citep{2008ApJ...673L.143M} 
with the Galactic Disk.  
Other HVCs may be tidal debris (e.g. \citealt{2003ApJ...591L..33L}, \citealt{2004ApJ...603L..77P}).  
Such coincidences yield distances to individual HVCs of 
order 10 kpc, but are quite rare and tell us little about the population as a 
whole.  Accurate distances to \HI\ clouds allow us to calculate their masses and 
to map their positions about the Galaxy and thereby differentiate between several 
HVC formation scenarios, which in turn have implications for cosmology and 
structure formation, chemical enrichment, and halo dynamics.

To explain the pervasive structure of Galactic HVCs across the sky, 
one must invoke multiple explanations.  The Magellanic Stream is likely the result 
of tidal interactions with the Large and Small Magellanic Clouds.  For other HVCs,
\citet{1976ApJ...205..762S} and \citet{1980ApJ...236..577B} postulated that a
galactic fountain mechanism could be responsible for raising hot gas into the
Galactic halo where it would cool, condense, and then rain back down on the
disk.  Meanwhile, \citet{1970A&A.....7..381O} proposed 
that these \HI\ clouds are primordial left-overs that are just now being
accreted by the Milky Way.  This theory was revived recently by 
\citet{1999ApJ...514..818B} and \citet{1999A&A...341..437B} who suggested that 
some types of HVCs were associated with low mass dark matter halos seen in 
models of $\Lambda$CDM galaxy formation \citep{1999ApJ...522...82K}.
HVCs may also be produced by a cooling, condensing hot halo 
(e.g. \citealt{2004MNRAS.355..694M}, \citealt{2006ApJ...639..590F}, \citealt{2006ApJ...644L...1S},
\citealt{2008ApJ...674..227P}). These theories have different predictions for the
mass distribution, the vertical distance from the disk, and the metallicity
for the clouds \citep[see][and references therein]{1997ARA&A..35..217W}.  But
in either case, their presence would not be unique to our own
Galaxy: in order to gain insight into the Galactic HVC population we
observe other nearby spiral galaxies.

As the closest large spiral galaxy to the Milky Way, M~31 is an excellent 
test bed in which to search for HVC analogs.  \citet{2004ApJ...601L..39T} find $\sim20$ 
discrete objects within 50 kpc of the M~31 disk which range in \HI\ mass 
between $10^5$ and $10^7$ \msun\ and can be characterized by a steep power
law distribution.  The authors suggest that these clouds may trace the dark
matter halos around M~31 that are predicted from $\Lambda$CDM models.
However, many of the clouds show velocity continuity implying that they may be
tidal remnants associated with the stellar stream of M~31, or part of the \HI\
bridge between M~31 and M~33 \citep{2005A&A...436..101W}.

Observing edge-on galaxies reveals the vertical distribution of gas.  The first, 
and subsequently, the most well studied galaxy for extraplanar \HI\ is the edge-on 
spiral NGC~891 (e.g. \citealt{1997ApJ...491..140S}).
The deepest published observations probe to \HI\ masses of $9.3\times10^{4}$
\msun\ and a column density of $1.6\times10^{18}$ cm$^{-2}$ \citep{2007AJ....134.1019O}.  
These observations show that NGC~891 has a large \HI\ halo that extends vertically 
out to 14 kpc from the thin \HI\ disk and, in one quadrant, out to 22 kpc, contributing 
nearly $30$\% of the total \HI\ mass of the galaxy.  This halo is characterized by
differential rotation, but it lags behind the rotation of the disk by $15$\kms\ kpc$^{-1}$.
Studies of NGC~891 and other nearly edge-on galaxies imply that there is a significant 
link between the amount of star formation in the disk (as measured by H$\alpha$ or radio 
continuum), and the properties of the lagging halo, which is believed to be evidence for gas 
circulation by a galactic fountain (\citealt{1997ApJ...491..140S}, \citealt{2006ApJ...636..181H,2006ApJ...647.1018H,2007ApJ...663..933H}).  In general, they find that the scale height and the velocity 
gradient decrease with increasing star formation rate.  Unfortunately, 
observations of edge-on galaxies are unable to tell us about either the position of
clouds around the disk, or their vertical motion---the latter of which plays a
part in identifying HVCs in the Milky Way.

In face-on galaxies \HI\ at anomalous velocities and \HI\ holes may be the 
result of gas accretion \citep[e.g. M~101;][]{1988AJ.....95.1354V} or massive 
star formation (e.g. NGC~628; \citealt{1992A&A...253..335K}).  The mass and 
morphology of clouds are frequently used to discriminate between high velocity complexes 
that are likely to be in-falling extragalactic gas versus galactic fountain gas
energized by supernovae.  In the case of accretion, the gas tends to be located
in the outer portions of the galaxy and is more massive.  In the star formation
case, the gas clouds tend to have lower masses and are associated with regions of 
H$\alpha$ or UV emission, or \HI\ holes
\citep[e.g.][]{2008arXiv0807.3339B}.  The largest \HI\ holes in M~101 may be the
result of recent collisions with extragalactic clouds; supernovae are ruled
out because of the energy requirement to account for the amount of \HI\ removed
from the disk, and there is no evidence from UV or radio continuum data to
support extraordinary star formation activity \citep{1988AJ.....95.1354V}.  
NGC~6946 has an \HI\ plume and large scale asymmetries probably related to 
tidal encounters \citep{1993A&A...273L..31K, 2008arXiv0807.3339B}, as well 
as high-velocity gas and holes related to star formation.  From the mass in the high
velocity gas, \citet{1993A&A...273L..31K} estimate a SN rate of 1 per 100 years
to feed a galactic fountain, which is roughly consistent with the six SNe that have been
observed in the galaxy this century.

Inclined galaxies have the benefits of both edge-on and face-on observations.
One can distinguish the presence of extraplanar \HI\ gas that lags behind the 
rotation of the disk and identify anomalous velocity gas with some sense of its location with
respect to the disk while, at the same time, look for sites of active star formation that
would be obscured in edge-on galaxies.  There are a number of good examples
of this.  Position-velocity (hereafter p-v) slices of NGC~2403 show \HI\ gas 
above and below the thin disk rotating $25-50$\kms\ slower \citep{2000A&A...356L..49S,2002AJ....123.3124F}, 
analogous to the extra-planar gas in NGC~891.  Furthermore, channel maps show
counter-rotating gas \citep{2001ApJ...562L..47F,2002AJ....123.3124F}.  In 
other examples, an asymmetry in the distribution of diffuse gas in NGC~253 is 
evidence for merger or accretion activity \citep{2005A&A...431...65B}.  
Meanwhile H$\alpha$ emission, from star formation activity, is correlated with 
high velocity gas in NGC~6822 \citep{2006AJ....131..363D}.  \HI\ holes have
been observed in NGC~4559 that are attributed to supernova explosions or 
stellar winds \citep{2005A&A...439..947B}.

In order to better understand the origin of HVCs in the Milky Way and
those seen around M~31, it is essential to observe a number of galaxies with
similar properties that are well-situated for detailed study.  NGC~2997 is 
the largest spiral galaxy in the loose galaxy group, LGG~180 \citep{1993yCat..41000047G}.  
This galaxy group was one of six observed by \citet{2007ApJ...662..959P} in a 
Parkes survey of nearby Local Group analogs to search for intra-group HVC candidates.  No
HVC analogs were found in any of the groups down to a mass limit $\lesssim10^7$ \msun.  
We chose NGC~2997 for deep, high resolution observations because of its similarities 
to the Milky Way and M~31 in mass, luminosity, and star formation rate.  It is a 
relatively isolated galaxy (no known companions within 110 kpc), it is moderately inclined 
with respect to the line of sight ($\sim33^{\circ}$), and it has with a well behaved 
rotation curve.  NGC~2997 was also chosen because it cannot be confused with 
Galactic or Local Group \HI\ emission.  Given a healthy star formation rate, it would not be surprising to find 
evidence for a lagging \HI\ disk; its inclination allows us to investigate 
the position and kinematics of clouds with respect to the galactic disk in 
simple rotation.  With these observations we identify extraplanar \HI\ and 
HVC analogs associated with galactic fountains, galaxy formation, and tidal debris
and use the characteristics of such HVCs to discriminate between their possible origins.

The observations of NGC~2997 and data reduction are described in Section~\ref{obs}.
The resulting data are described in 
Section~\ref{results} including the results from modelling each \HI\ component 
in NGC~2997.  In Section~\ref{discussion} we consider possible origins of 
the extraplanar gas of the thick disk and the anomalous velocity \HI\ clouds, 
and compare the \HI\ in NGC~2997 to that of other galaxies.  We also derive the
star formation rate and distance of NGC~2997, and discuss how it affects the 
constraints on the origins of HVCs.  We conclude in Section~\ref{conclusion}.

\section{Observations \& Data Reduction}
\label{obs}

In January 2006, we observed NGC~2997 with the Australia Telescope Compact
Array (ATCA) for 28 hours in the EW682 configuration.  These observations were
in addition to archival ATCA data (from project C453) for NGC 2997 taken during
1995 and 1996, amounting to a total of 59
hours in 6 different configurations (Table~\ref{ATCAobs}).  The data were 
flagged and calibrated using the standard procedures in Miriad 
\citep{2008...Miriad}.  From 15-26 January 2007, we observed the galaxy for a 
total of 61 hours in 7-8 hour blocks with the Giant Metrewave Radio Telescope 
(GMRT).  The GMRT is comprised of 30 antennas in a fixed ``Y''-configuration 
with 14 antennas within 1 km and a maximum baseline of 25 km.
The data were flagged and calibrated in AIPS at the GMRT using the standard 
procedures. Each night was individually imaged and inspected before being combined to
create our final data cube.  All observations were taken at a velocity
resolution of 6.596\kms.  
 
To combine the data, the reduced GMRT observations were written out of
AIPS and read into Miriad \footnote{In Miriad, the imported image headers
  were incomplete: we had to set the rest frequency to 1.420405752 GHz and,
  for the purposes of estimating the rms noise before imaging, we added the
  correct values for the system temperature (76 K) and system gain (0.22
  Jy/K/antenna $\times$ 30 antennas).
  \url{http://www.gmrt.ncra.tifr.res.in/gmrt\_hpage/Users/doc/spec\_mod.pdf}}.
The visibility data were combined in Miriad with the task UVAVER, and cleaned 
using the task MOSSDI.  We imaged 120 channels centered on the galaxy using a 
robustness of 2 and down-weighted
the longest baselines from the GMRT.  
We derive a distance to NGC~2997 of 12.2 Mpc using published 2MASS Tully-Fisher 
relations \citep{2008AJ....135.1738M} that we adopt throughout the paper.  
Discussion of this derivation is left to the Section~\ref{distance} in order to not 
take away from the focus of this paper on the H~I properties of the galaxy.  At this distance, our 
observations have a spatial resolution of 14\arcsec\ = 0.8 kpc.  Altogether, 
our final cube has a rms noise of $\sigma = 0.404$ mJy\kms\ 
beam$^{-1}$ channel$^{-1}$.  At 5$\sigma$ we achieve a mass sensitivity of $4.7\times10^5$ 
\msun\ channel$^{-1}$ and a column density of $9.6\times10^{19}$ 
cm$^{-2}$ beam$^{-1}$ channel$^{-1}$.

\begin{table}[t]
  \begin{center}
    \caption{Summary of ATCA Observations \label{ATCAobs}}
    \begin{tabular}{ccc}
      \tableline\tableline
      & Configuration/ & Observing \\
      Date & Max Baseline (m)\tablenotemark{a} & Time (hrs) \\
      \tableline
      1995 Oct 27 & 1.5D/1439 (4439) & 10 \\
      1995 Nov 7 & 6A/2923 (5939) & 14 \\
      1995 Dec 15 & 6C/2786 (6000) & 13 \\
      1996 Jan 21 & 750C/750 (5020) & 12 \\
      1996 Jan 23 & 750B/765 (4500) & 11 \\
      1996 Feb 22 & 1.5C/1485 (4500) & 9 \\
      2006 Jan 7 & EW352/352 (4439) & 8.5 \\
      2006 Jan 8 & EW352/352 (4439) & 8.5 \\
      2006 Jan 9 & EW352/352 (4439) & 11 \\
      \tableline
    \end{tabular}
    \tablenotetext{a}{Maximum separation between antennas, excluding (and including)
the 6 km antenna.}
  \end{center}
\end{table}

\section{Results}
\label{results}

\begin{figure*}[t]
\centering
\includegraphics[width=0.9\textwidth]{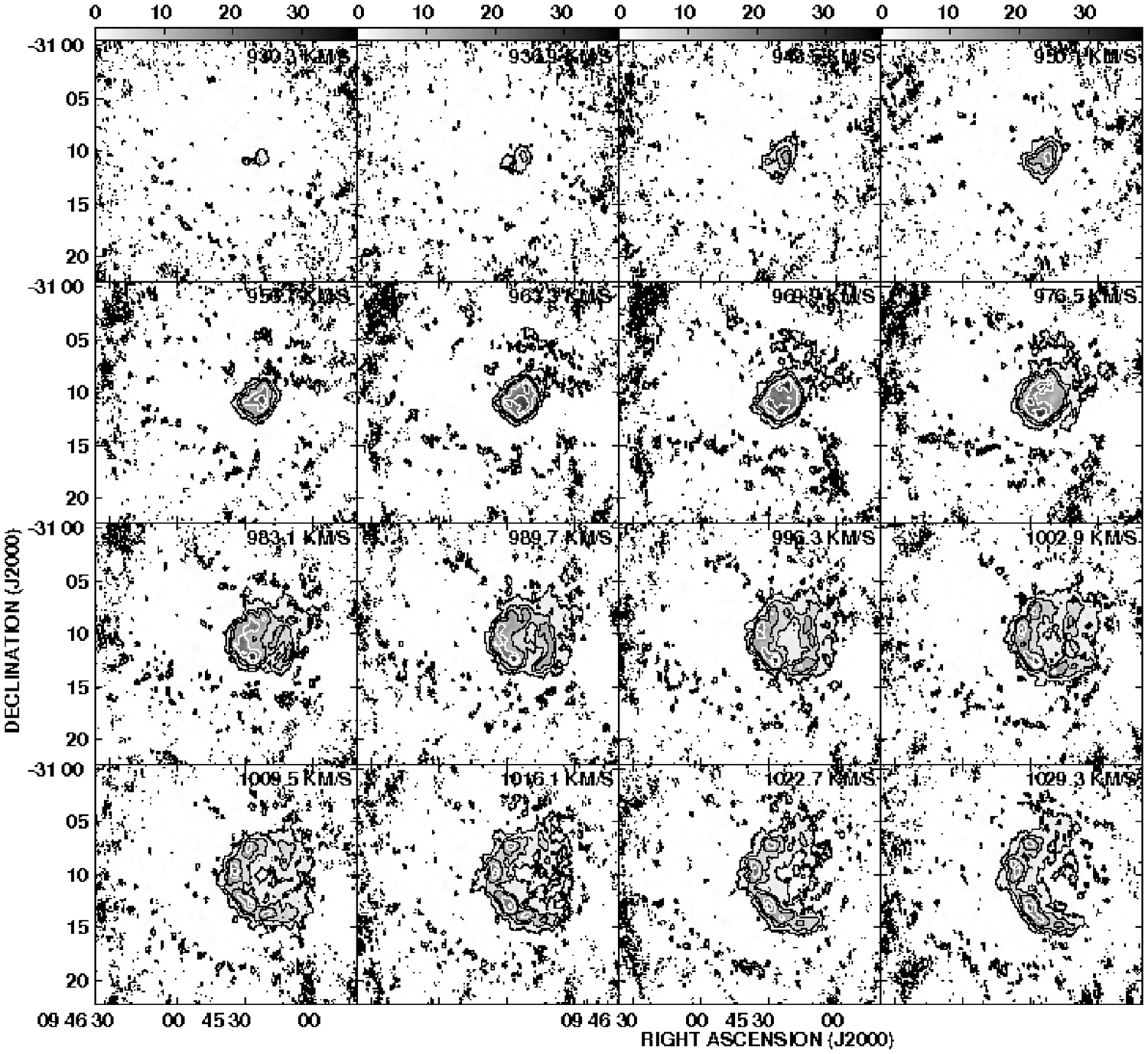}
\caption{Channel maps of NGC~2997 covering 930.3 to 1029.3\kms\ 
heliocentric velocity. Data is plotted at full spatial and spectral 
resolution. Contours are plotted for 3, 10, 20, 40, 60$\sigma$, where 
$\sigma = 1.9\times10^{19}$ cm$^{-2}$ beam$^{-1}$ channel$^{-1}$.
Grey scale is in units of mJy beam$^{-1}$.} 
\label{chan1}
\end{figure*}

\begin{figure*}[t]
\centering
\includegraphics[width=0.9\textwidth]{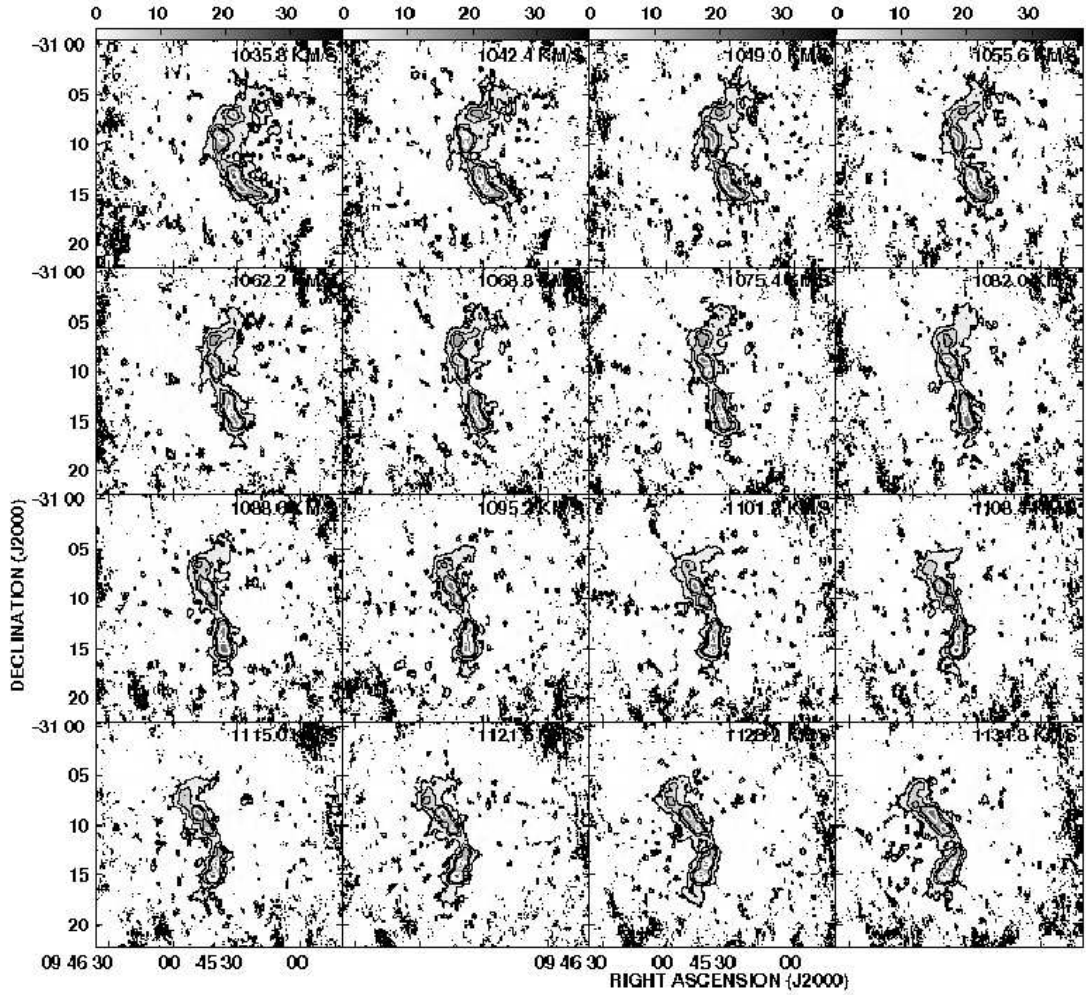}
\caption{As in Figure~\ref{chan1} but covering 1035.8 to 1134.8\kms.}
\label{chan2}
\end{figure*}

\begin{figure*}[t]
\centering
\includegraphics[width=0.9\textwidth]{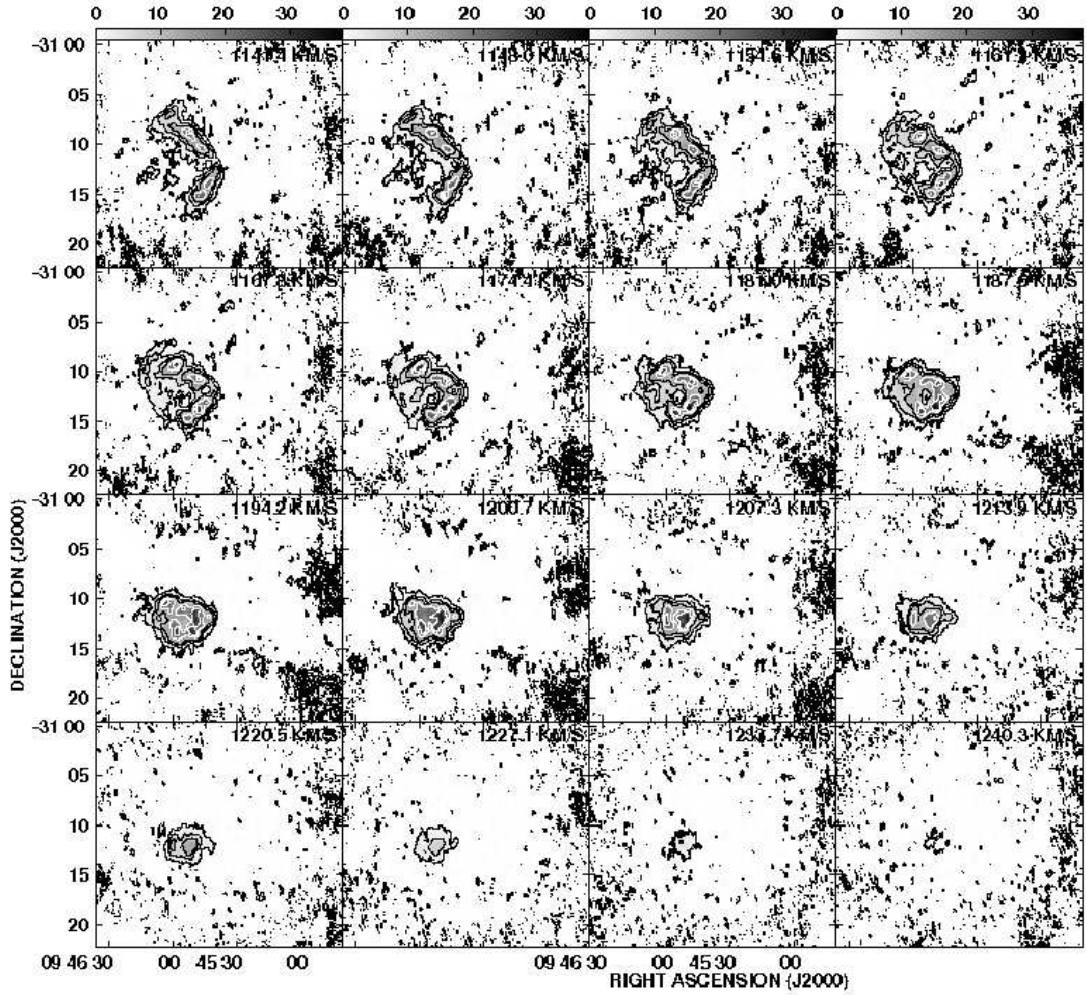}
\caption{As in Figure~\ref{chan1} but covering 1141.3 to 1240.3\kms.}
\label{chan3}
\end{figure*}

We present our data cube in the form of channel maps in Figures~\ref{chan1}-\ref{chan3}.
These channel maps have been clipped at the 3$\sigma$ level to emphasize the 
galactic emission.  The figures illustrate the general rotation 
of NGC~2997, but also show evidence of a thick disk and additional \HI\ that 
is moving at velocities inconsistent with a rotating thin disk.  

\begin{figure}[t!]
\centering
  \includegraphics[width=0.9\textwidth]{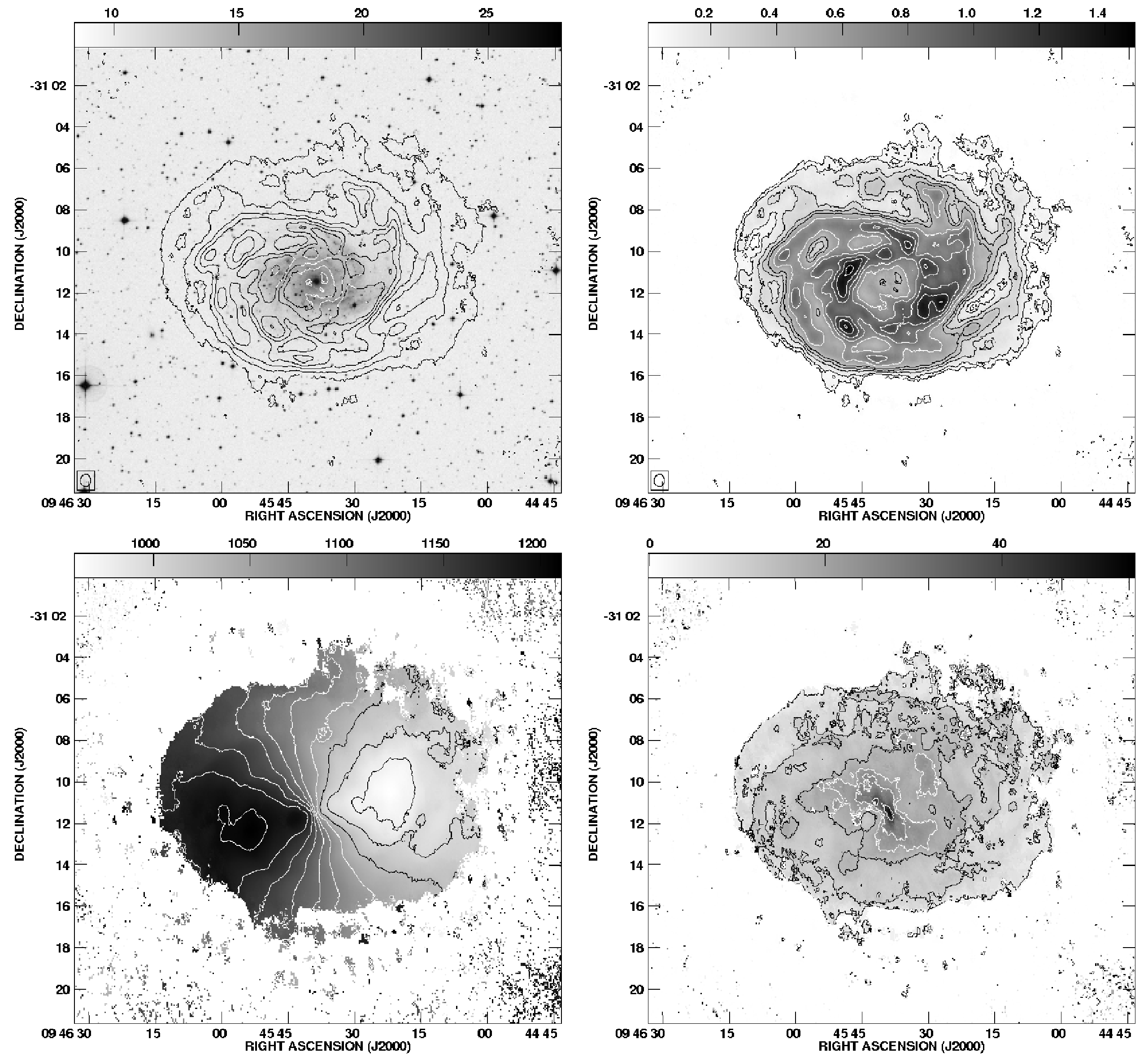}
   \caption{Top Left:  DSS image of NGC~2997 overlayed with \HI\ contours.
       Top Right:  \HI\ total intensity map; contours are 
       at 5, 20, 50, 80, 120$\sigma$, where $\sigma = 1.9\times10^{19}$ 
       cm$^{-2}$ beam$^{-1}$ channel$^{-1}$.  Grey scale is in units of Jy beam$^{-1}$.
       Bottom Left:  Intensity weighted velocity map of NGC~2997; contours range 
       over $975-1200$\kms\ and are spaced every 25\kms.  
       Bottom Right:  the intensity weighted velocity dispersion map; contours 
       are at 4, 10, 15, 20, 25, 40\kms.}
  \label{moment}
\end{figure}

In Figure \ref{moment} we present moment maps created from the data cubes: the optical
DSS image with \HI\ contours, the total \HI\ intensity map, the velocity field, and the
velocity dispersion (moments 0-2).  The moments are calculated independently for
each pixel within a specified clip range---in our case, it is calculated from
pixels containing emission above 5$\sigma$ of our rms sensitivity in order to avoid
large outliers, particularly on the outskirts of the galaxy where the noise is increasing with 
distance from the pointing center.  In the optical image, it becomes evident that the \HI\
gas is extended to the northwest with respect to the optical disk.  However, the intensity
weighted velocity map looks fairly well behaved; there is no direct evidence
that the galaxy has been disturbed by a strong encounter.  Finally, the velocity dispersion 
map shows that the greatest dispersion occurs in the center likely due to beam smearing 
effects and due to intense star formation as suggested by strong radio continuum 
emission at the center of the galaxy.

\begin{figure}[h]
\centering
\includegraphics[width=0.9\textwidth]{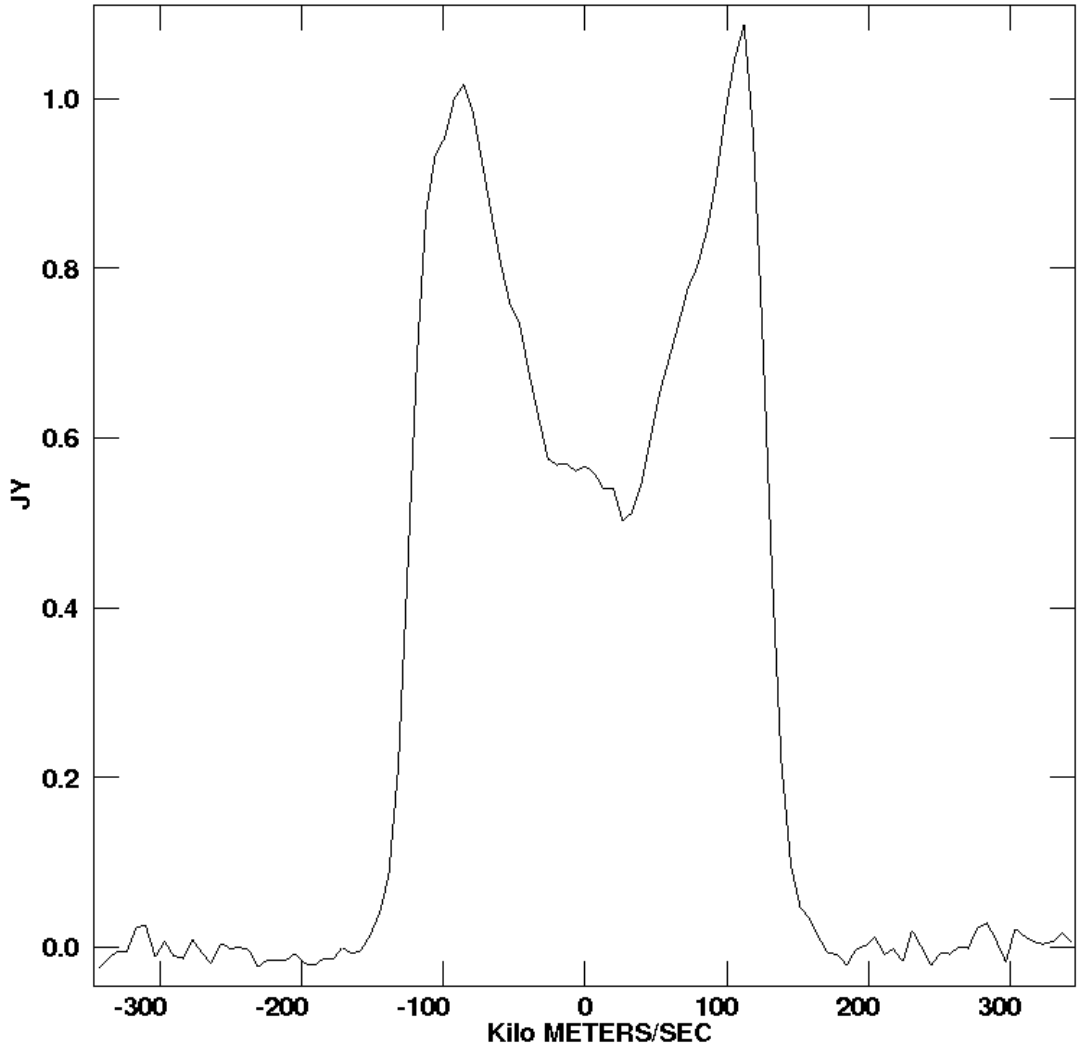}
\caption{The integrated \HI\ profile of NGC~2997 over the full \HI\ extent of
the galaxy. The profile is centered on the systemic velocity of the galaxy, 1089\kms.}
\label{profile}
\end{figure}

Figure \ref{profile} shows the integrated \HI\ profile over the
area of the entire galaxy.  Summing the emission in 
this profile yields a total flux for NGC~2997 of 226.6 Jy\kms.  This 
compares to 162.3 Jy\kms\ from HIPASS \citep{2004MNRAS.350.1195M}.  However, the 
latter value is likely to be low since the HIPASS measurement assumed the 
galaxy was unresolved.  At a distance of 12.2 Mpc, our measured flux
corresponds to a total \HI\ mass of $7.96\times10^{9}$ \msun.   

\begin{figure}
\centering
\includegraphics[width=0.4\textwidth]{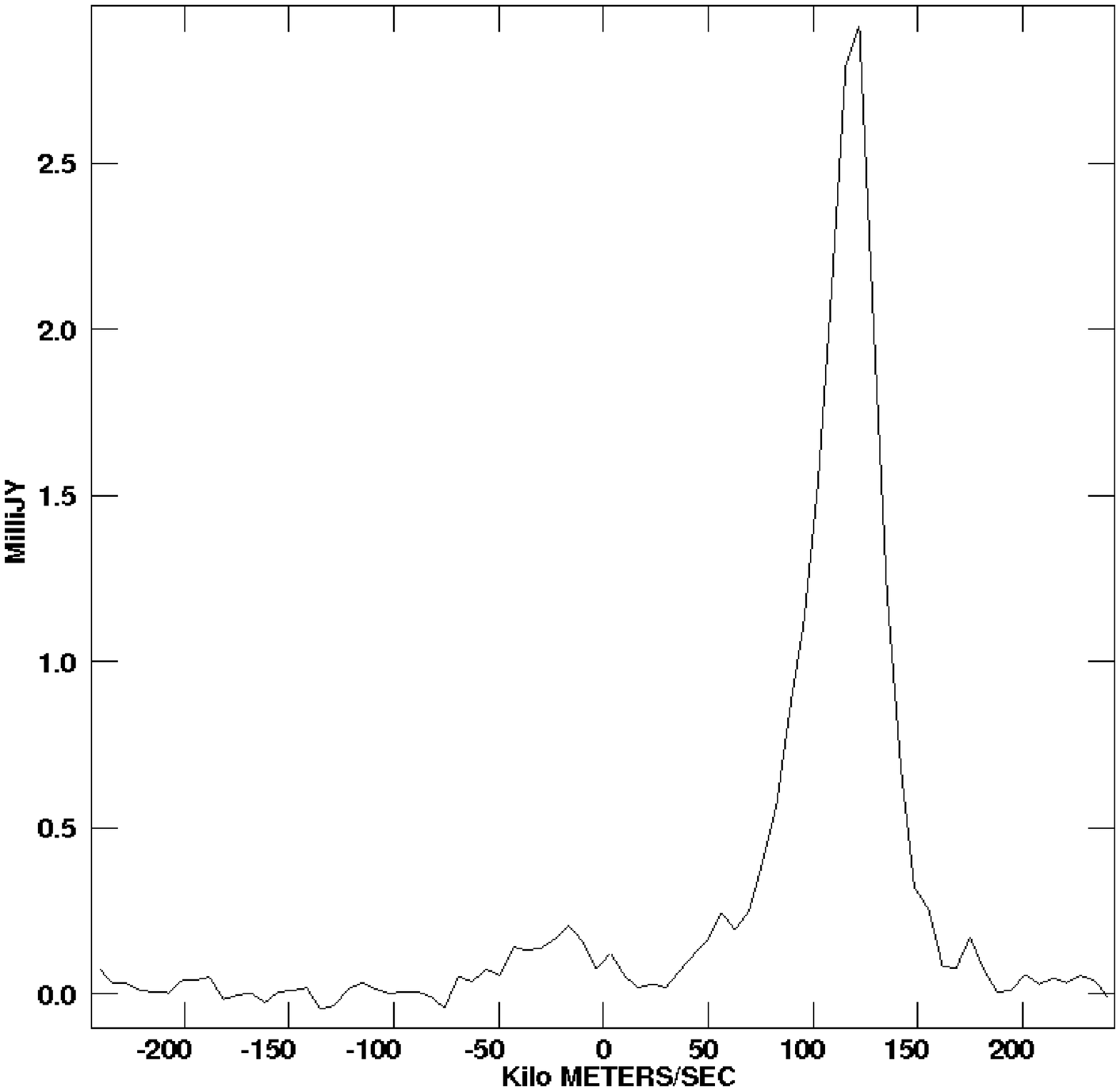}\hfill
\includegraphics[width=0.4\textwidth]{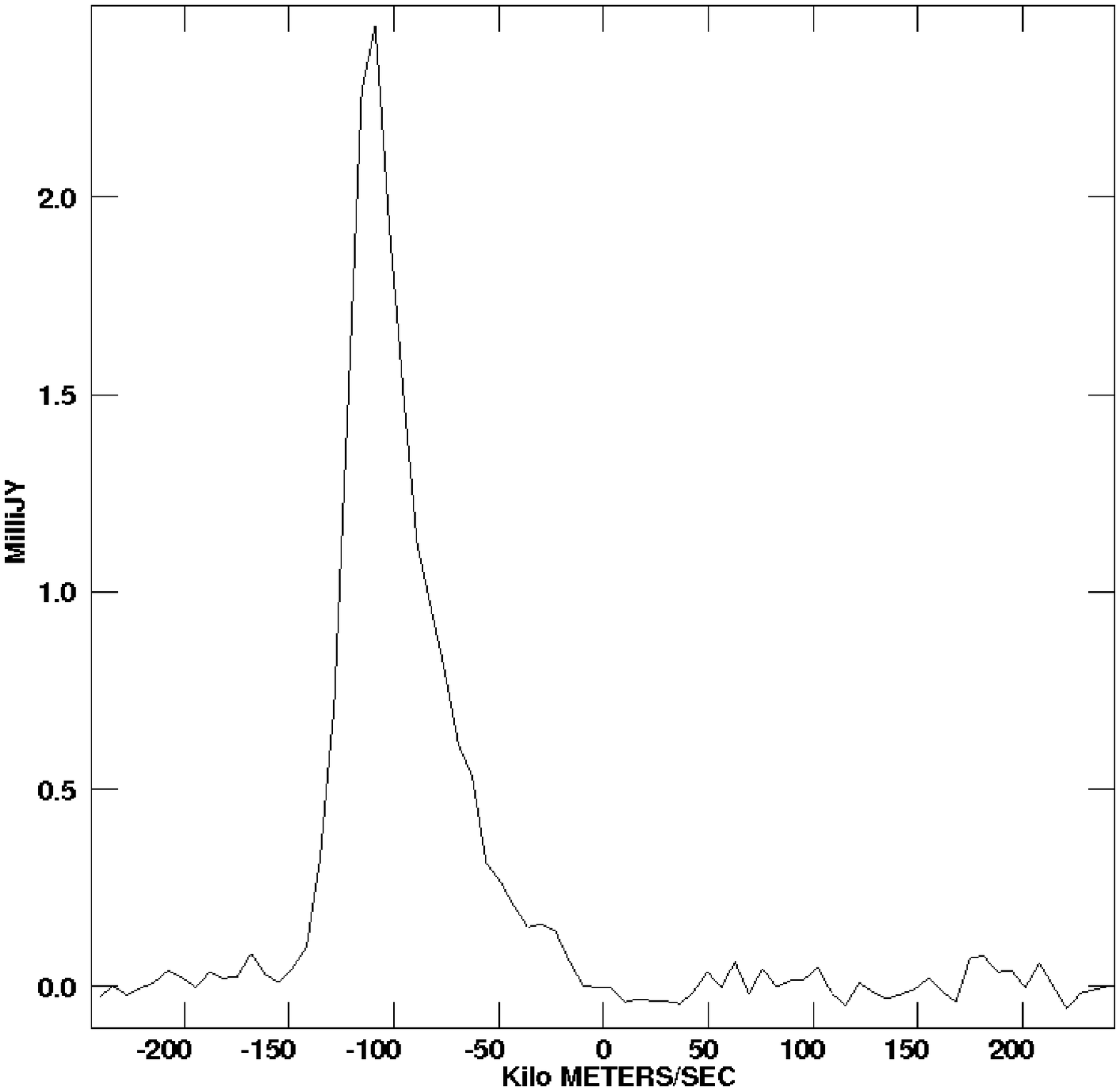}
\caption{Left:  A sample spectrum towards a position on the eastern side of NGC~2997 
showing the thin disk rotation at $\sim120$\kms; the asymmetry of the \HI\ profile, 
characteristic of the thick disk, at $\sim60$\kms; and 
anomalous \HI\ at $\sim-25$\kms.  Right:  A spectrum from the western side of the galaxy
illustrating the thin ($\sim110$\kms) and thick disk ($\sim50$\kms) components.  The spectra
are centered on the systemic velocity of the galaxy, 1089\kms.}
\label{examplespectra1}
\end{figure}

We are interested in understanding the nature of the \HI\ in NGC~2997 and
identifying gas clouds that are analogous to the HVCs observed
in the Milky Way.  In decomposing the galaxy, we find there are three
components that contribute to its total \HI\ content.  A thin disk dominates the 
total \HI\ emission and can be modeled by a Gaussian profile with a velocity 
dispersion of $11\pm3$\kms.  We modeled and fit a rotation curve to this thin 
disk, however, there is additional gas that appears as a wing on the side of 
the thin disk's velocity profile towards the systemic velocity.  This 
``thick disk" contributes approximately $16-17$\% of the total \HI\ mass and 
has a velocity dispersion of $\sim16$\kms.  While this second component shows 
regular rotation, it lags behind the thin disk at a relatively constant rate.  
The third component consists of \HI\ emission that cannot be explained in terms 
of either disk component, and some of which is counter-rotating.
These three components stand out in Figure \ref{examplespectra1}.

Table \ref{properties} summarizes the properties we derive for the galaxy.  
We calculate a new distance to NGC~2997 based on the 2MASS Tully-Fisher relations 
from \citet{2008AJ....135.1738M}.  The radio center, systemic velocity, inclination, 
position angle, and the maximum rotational velocity come from our rotation curve 
fit to the velocity field. The integrated flux and \mhi\ come directly from the data cube.  
$M_{dyn}$ is calculated from the maximum rotational velocity.  The velocity gradient,
scale height and cyclindrical height come from our best fit model of the extraplanar 
\HI.  The star formation rate is estimated from values in the literature.  Sections 
\ref{thindisk}-\ref{anomaloushi} and Discussion describe in detail how 
we arrive at these values.

\begin{table}[h]
  \begin{center}
    \caption{Properties of NGC~2997 \label{properties}}
    \begin{tabular}{ll}
      \tableline\tableline
      Property & Value \\
      \tableline
      Hubble Type & SAc\tablenotemark{a} \\
      Optical Center (J2000) & 09$^h$45$^m$38.8$^s$ -31$^{\circ}$11$^\prime$27.9\arcsec\tablenotemark{a} \\
      Optical Extent & 8.9\arcmin $\times$ 6.8\arcmin\tablenotemark{a} \\
      Distance & 12.2 Mpc\tablenotemark{b} \\
      Radio Center (J2000) & 09$^h$45$^m$35.9$^s$ -31$^{\circ}$11$^\prime$27.8\arcsec \\
      Radio Extent & 15.6\arcmin $\times$ 11.1\arcmin\tablenotemark{c} \\
      Systemic Velocity & $1089\pm5$\kms \\
      Inclination, \it{i} & $32.3^{\circ}\pm0.3$ \\
      Postion Angle & $107.1^{\circ}\pm0.6$ \\
      Integrated Flux Density & 226.6 Jy\kms \\
      \HI\ Mass, \mhi & $7.96\times10^{9}$ \msun\tablenotemark{b} \\
      Maximum $v_{rot}$ & $226\pm2$\kms \\
      Dynamical Mass, $M_{dyn}$ & $2.1\times10^{11}$ \msun \\
      Velocity gradient, $\frac{dv}{dz}$ & $18-31$\kms\ kpc$^{-1}$ \\
      Scale Height, $z$ & $0.7\pm0.2$ kpc \\
      Cyclindrical Height, $z_{cyl}$ & $0.5\pm0.1$ kpc \\
      Star Formation Rate & $5\pm1$ \msun\ yr$^{-1}$ \\
      \tableline
    \end{tabular}
    \tablecomments{Unless otherwise stated, the values in Table \ref{properties} are derived from this work.}
    \tablenotetext{a}{Optical position and extent obtained from the NASA/IPAC 
      Extragalactic Database (NED).}
    \tablenotetext{b}{Distance and \HI\ mass are derived from the 2MASS Tully Fisher 
      relations of \citet{2008AJ....135.1738M}.}
    \tablenotetext{c}{Measured from 3$\sigma$ contours along major and minor axes.}
  \end{center}
\end{table} 

\subsection{Modeling the Thin Disk}
\label{thindisk}

\begin{figure}[t!]
\includegraphics[height=0.8\textheight]{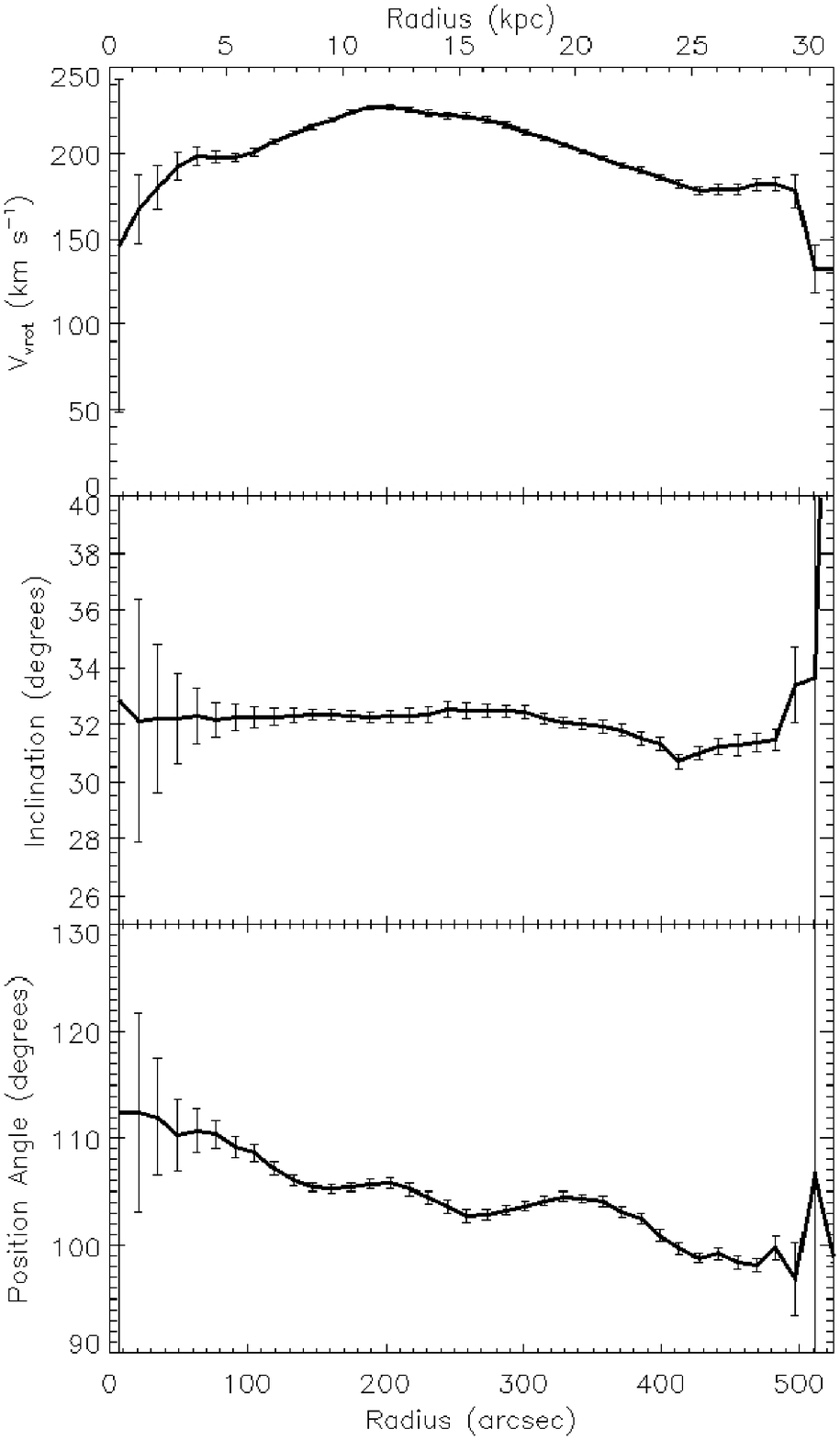}
\caption{The top panel is our rotation curve derived from a tilted ring model.
It peaks at 226\kms\ around $12$ kpc.  The two bottom panels
  demonstrate the quality of our assumption that NGC~2997 is an ideal, flat
  disk.  We keep all other parameters fixed but allow the inclination 
  (middle) and position angle (bottom) to vary---the rotational velocity for each tilted ring 
  is fixed to the value found in the top panel.  The declining position angle with radius
  suggests we were unable to ft a self-consistent rotation curve.}
\label{rotcur}
\end{figure}

As illustrated by Figure~\ref{examplespectra1}, the \HI\ profile towards most
positions in NGC 2997 is dominated by a narrow velocity component with
a dispersion of $11\pm3$\kms.  To model it, we clipped the data cube at 10\% of the
galaxy's peak emission \citep[in a manner similar to][]{2002AJ....123.3124F} and 
fit a Gaussian to this upper part of the velocity profiles.  By clipping the 
emission at this level we prevent the subsequent analysis of NGC~2997's thin 
disk from being biased by broad, low column density
\HI\ velocity components.  We fit a rotation curve to the thin disk following the 
procedure of \citet{1989A&A...223...47B}.  We assume that the position angle 
and inclination of the galaxy are constant for all radii (we assume an ideal, 
flat disk), and from this we derive a systemic velocity of $1089\pm5$\kms, a 
position angle of $107.1^{\circ}\pm0.6$ and an inclination of $32.3^{\circ}\pm0.3$.  
The errors are the average errors found for rings between $5-30$ kpc, however, Figure
\ref{rotcur} shows that there is a gradient in the position angle, which has not been
taken into account.

The top panel of Figure \ref{rotcur} shows our best fit rotation curve.  It
rises out to $\sim12$ kpc, then slowly declines to greater than 25 kpc.
We trust this rotation curve to $\sim28$ kpc, beyond which the lopsidedness 
of the \HI\ emission produces large errors in the fit.  In addition, in order to understand 
the quality of our assumption that the disk is flat and ideal, we re-fit the
rotation curve allowing the inclination and the position angle
to vary while keeping the rotational velocity fixed to the value found for each ring
in the tilted ring model.  The middle and bottom panels of Figure \ref{rotcur} show how the
inclination and position angle, respectively, vary with radius.  In these
instances the rotational velocity for each ring is fixed to the value plotted in
the top panel.

There is evidence that NGC~2997 deviates from the ideal, flat disk.  
The bottom panel of Figure
\ref{rotcur} suggests that we were unable to fit a self-consistent rotation
curve with a uniform position angle because it declines steadily with radius.
In the tilted-ring model, inclination and rotation velocity are coupled
through a $v_{rot}\sin i$ term.  This makes for a degeneracy between a flat
inclination and falling rotation curve that is particularly strong for low
inclination galaxies.  The ``S'' like appearance of contours near the minor
axis (Figure \ref{moment}) indicate that NGC~2997 may have a 
slightly warped \HI\ disk.

\subsection{Modeling the Lagging Thick Disk}
\label{thickdisk}

Our deep observations of NGC~2997 show that the \HI\ gas is not confined to
the thin disk.  In p-v slices the contours of
\HI\ emission show a shallow gradient towards the systemic velocity: a feature
which has been identified in a small number of other galaxies and has been
referred to as a ``beard'' \citep[e.g. NGC 2403;][]{2000A&A...356L..49S,2002AJ....123.3124F}.
After modeling the dominant \HI\ component, we subtract the thin disk
from the original data cube and find that $1.4\times10^9$ \msun\ of residual \HI\
emission remains---about 18\% of the total \mhi\ of the galaxy.

\begin{figure}[t]
\centering
\includegraphics[width=0.9\textwidth]{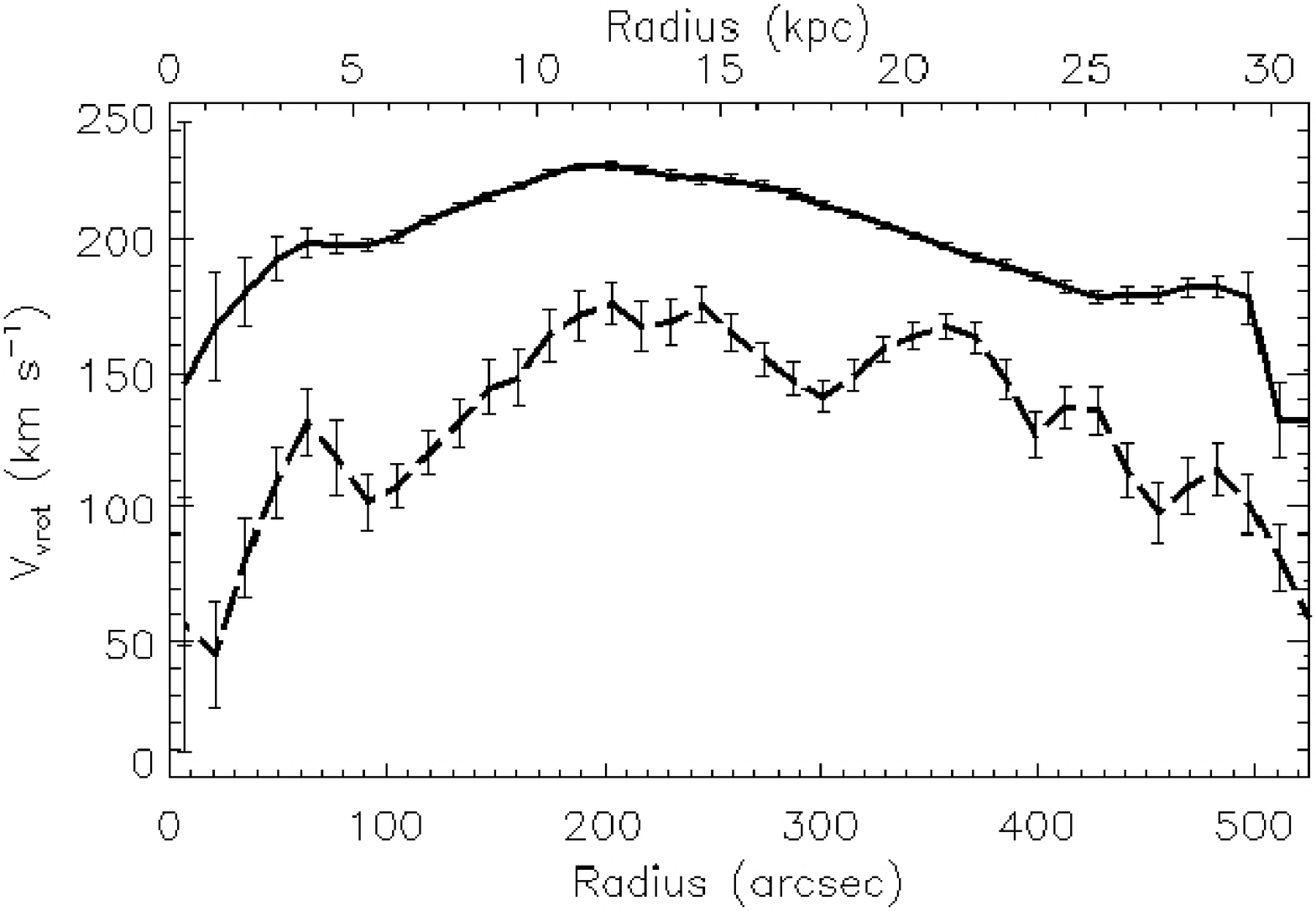}
\caption{Rotation curves for both the thin disk (solid line) and the thick disk
  (dashed line).  The average $\Delta v$ between the rotation curves is about $60$\kms.}
\label{rotcur2}
\end{figure}

This second component can be fit by a Gaussian with a higher velocity dispersion 
than the thin disk ($\sim16$\kms) which reflects
the thermal motions of the gas convolved with a vertical velocity gradient
\citep{2006AJ....131..363D}.  The peak velocity of the fitted Gaussian
resides closer to the systemic velocity of the galaxy than gas in the thin
disk at the same radius.  In other words, it appears to follow
galactic rotation, but at slower rotational velocities.  We fit a second
rotation curve to the residual data by assuming that this thick disk has the
same large-scale morphology as the thin disk, with the same systemic velocity
(1089\kms), position angle (107$^{\circ}$) and inclination
(32$^{\circ}$).  From Figure \ref{rotcur2} we see that peaks
in the two rotation curves roughly coincide with one another, implying that
they are related, however the thick disk consistently lags behind that of 
the thin disk by $\sim60$\kms.  A similar lagging disk has been observed
in NGC~891 ($25-200$\kms; \citealt{1997ApJ...491..140S}) and NGC~2403 
($25-50$\kms; \citealt{2000A&A...356L..49S,2002AJ....123.3124F}).

To understand this relation we also attempted to derive the thick disk 
rotation curve from the residual data without assuming it has the same 
properties as the thin disk.  While we successfully fit a position
angle ($\sim25^{\circ}$), we were unable to find a consistent inclination.
In NGC~2403, \citet{2002AJ....123.3124F} show that a rotated position angle,
such as this can be explained by an overall inflow of the anomalous gas.  However, 
as we find no other evidence for radial gas motion when we include it in our 
models of NGC~2997, we disregard it.

\begin{figure}[t]
\centering
\includegraphics[width=0.9\textwidth]{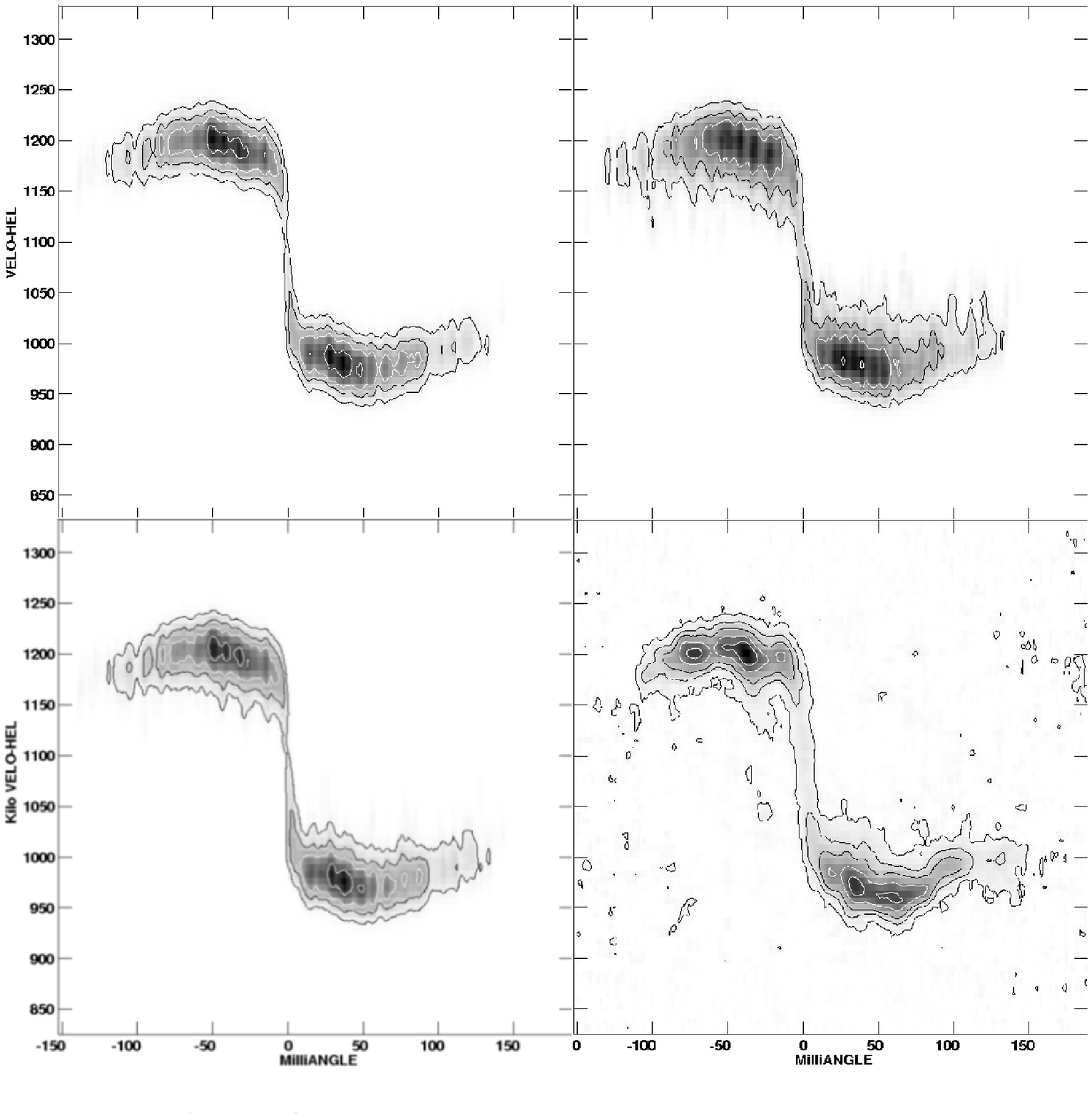}
\caption{Limiting examples of our lagging disk models.  Upper left-- the best 
scale height and $z_{cyl}$, but too small a velocity gradient: there is not 
enough emission close to the systemic velocity.  Upper right-- the best 
velocity gradient and $z_{cyl}$, but no scale height: the emission is more 
spread out in velocity space, and the peak is lower in intensity and closer to 
the systemic velocity than seen in the data.  Lower left-- our best fit model 
with scale height = 0.8 kpc, $z_{cyl} = 0.6$ kpc, and velocity gradient = 38\kms\ kpc$^{-1}$.
Lower right-- the data.  Contours are at 3, 10, 20, 40, 60$\sigma$, where $\sigma=1.9\times10^{19}$ 
cm$^{-2}$ beam$^{-1}$ channel$^{-1}$.}
\label{models}
\end{figure}

We took this kinematic modeling a step further by adding a velocity
gradient to the rotation curve as a function of height above the disk
mid-plane.  We followed similar procedures to those demonstrated by 
\citet{2006ApJ...636..181H} and \citet{2007AJ....134.1019O}.
The GIPSY task GALMOD allows one to create a model data cube from user
supplied parameters: the rotational velocity, the scale height and the surface
density.  A modified version of this task (Heald, 2008, private communication) 
includes two additional parameters to specify (1) the rate of change in 
velocity, $dv/dz$, with height, $z$, and (2) a cylindrical height 
above the disk at which the gradient begins, $z_{cyl}$, such that the rotation
of the gas is described by the following equations:
\begin{eqnarray}
  \label{equ:halomodgalz}
  v_{rot}(R,z)=\left\{
    \begin{array}{lll} v_{rot}(R,0) &  z \le z_{cyl},\\
      v_{rot}(R,0)-(z-z_{cyl})\frac{dv}{dz}  &  z \ge z_{cyl}
    \end{array}\right\}.
\end{eqnarray}
We examined p-v slices along the galaxy's major axis to compare the models 
created from the modified GALMOD task with our data (Figure \ref{models}).

The cylindrical height was the easiest parameter to fit.  It is defined as the distance 
from the mid-plane at which the velocity gradient begins.  Setting its value has the
effect of imposing the velocity at which the peak of the emission occurs.  For example, 
with a smaller cylindrical height the velocity gradient begins closer to the mid-plane 
and, for an inclined galaxy, the resulting sum of emission projected along 
the line of sight appears to shift towards lower velocities.
A cylindrical height of $0.5\pm0.1$ kpc fits our data best.
This is within the range of values derived for other galaxies:
\citet{2007ApJ...663..933H} report values ranging from 0.4 kpc in NGC~4302 to
1.2 kpc in NGC~5775 using H$\alpha$ kinematics of ionized gas; while 
\citet{2007AJ....134.1019O} report a value of 1.3 kpc for NGC~891 in \HI.  

The value of the scale height predominately influences the lowest level 
\HI\ contours of the p-v diagram.  
A large scale height places more emitting material at greater
distances from the disk and, therefore, at lower velocities with respect to
the rotational velocity at the mid-plane so the \HI\ contours become shallower
toward the systemic velocity with increasing scale height.  
The best fit for the scale height is $0.7\pm0.2$ kpc.  Our errors are based
on the resolution of parameter space sampled and the degree to which the models differed
significantly by eye.

The model is least sensitive to the velocity gradient, probably due to
NGC~2997's relatively low inclination and the low column density of the gas 
in the lagging disk.  Varying this value changes the
gradient of the \HI\ contours in the model, however it does so less
dramatically than varying the scale height.  This, combined with
the clumpiness of the \HI\ due to spiral arms and other features in the disk,
make it difficult to estimate a good fit.  We estimate the velocity gradient
to be between $18-31$\kms\ kpc$^{-1}$.  Again, these values fall in the
range reported for NGC~4302 ($22.8-59.2$\kms\ kpc$^{-1}$;
\citealt{2007ApJ...663..933H}), NGC~5775 (8\kms\ kpc$^{-1}$;
\citealt{2006ApJ...636..181H}), and NGC~891 (15\kms\ kpc$^{-1}$;
\citealt{2007AJ....134.1019O}).

\subsection{Anomalous \HI}
\label{anomaloushi}

\begin{figure}[h]
\includegraphics[width=0.9\textwidth]{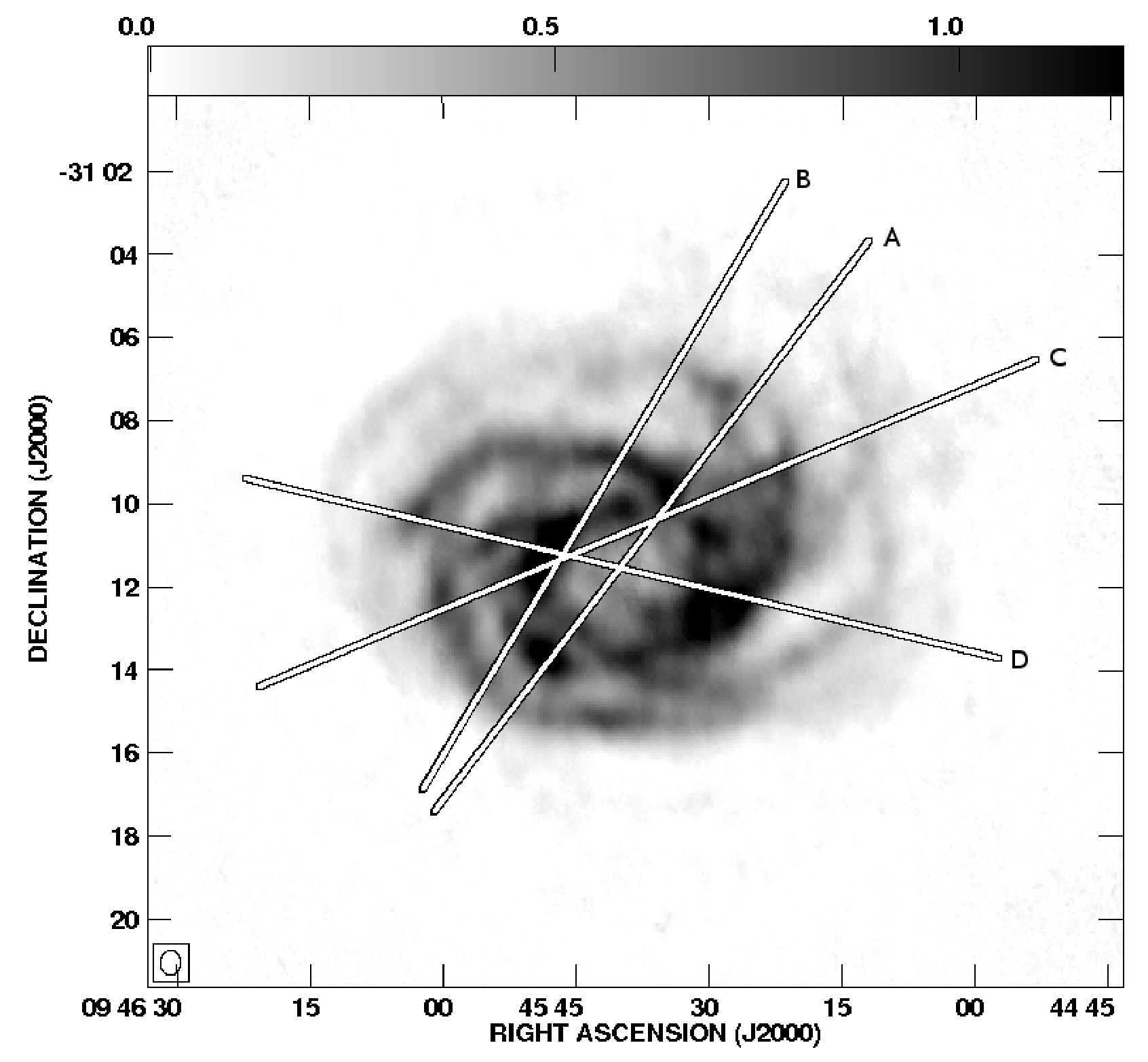}
\caption{A total intensity \HI\ map of NGC 2997 illustrating where slices have
  been made to analyze the anomalous \HI\ content (see Figure \ref{allslices}).}
\label{mom0}
\end{figure}

\begin{figure}[t]
  \includegraphics[width=1\textwidth]{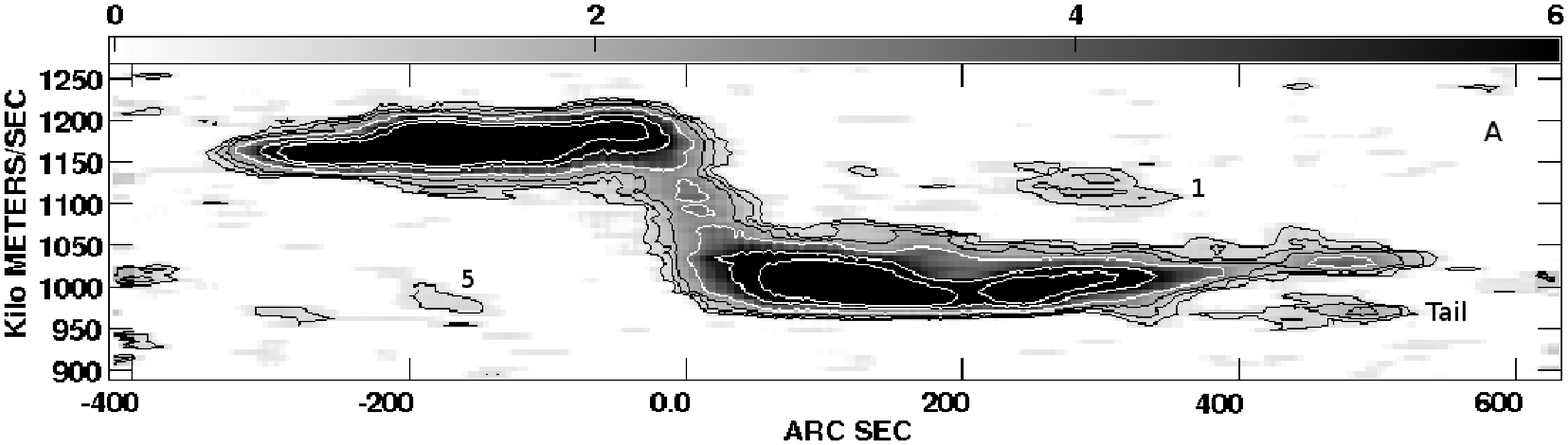}
  \includegraphics[width=1\textwidth]{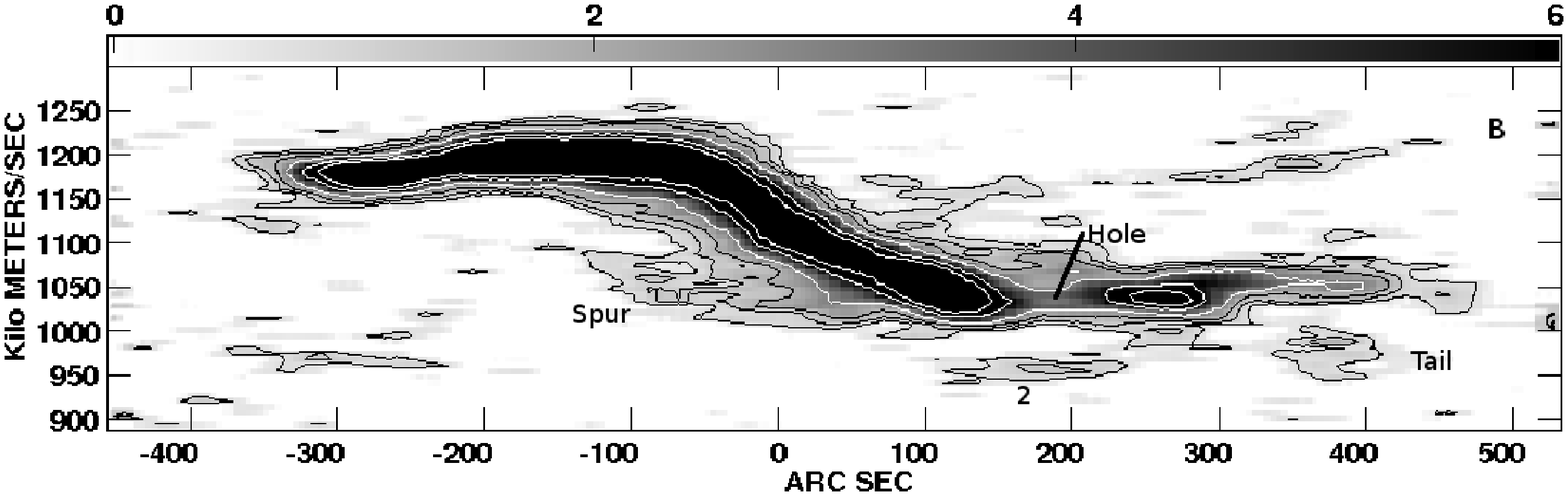}
  \includegraphics[width=1\textwidth]{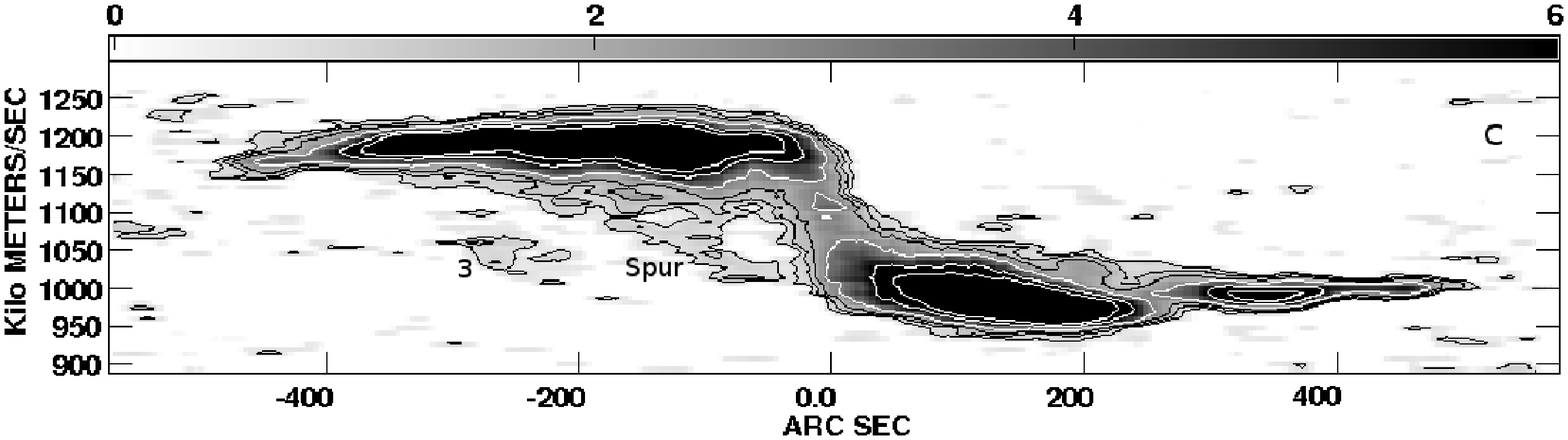}
  \includegraphics[width=1\textwidth]{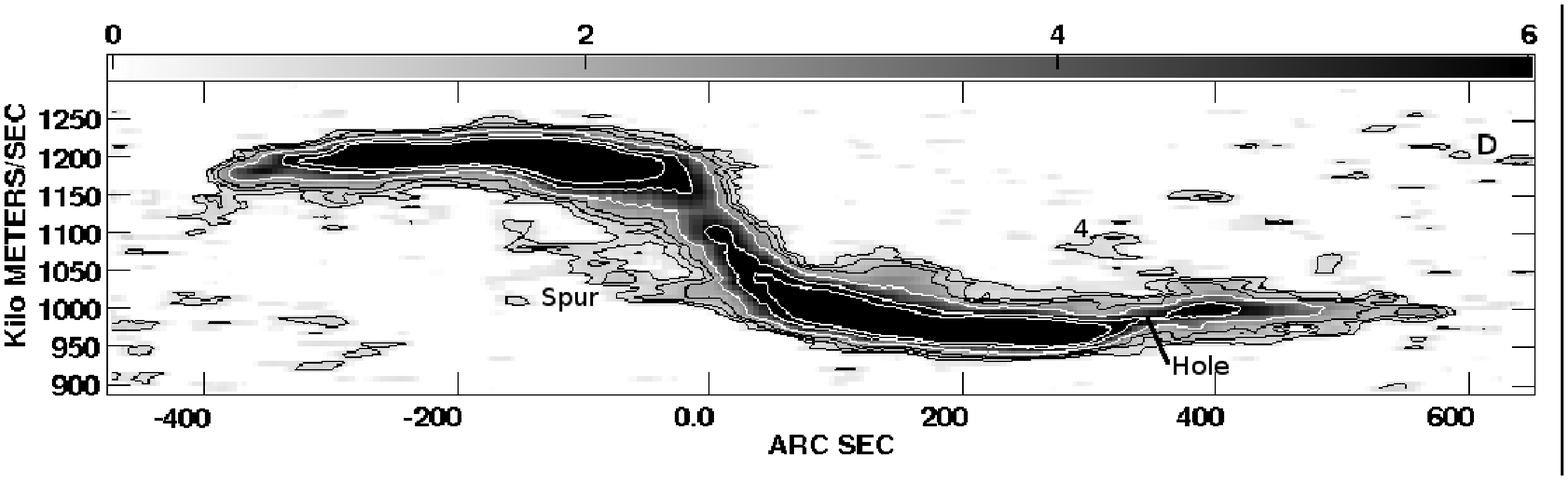}
  \caption{Select slices through the smoothed \HI\ cube illustrate the extraplanar \HI\ 
   (see Figure~\ref{mom0} \& Table~\ref{cloudprops}). (A) Clouds 1, 5, and the 
   \HI\ tail. (B) Cloud 2 with \HI\ hole, the \HI\ tail, spur, and additional 
   examples of faint, counter-rotating gas. (C) Cloud 3 and \HI\ spur. (D) Cloud 
   4 with \HI\ hole, and \HI\ spur.  The positive direction corresponds to the western 
   side of the galaxy.}
  \label{allslices}
\end{figure}

In spite of this extensive modeling, there is a significant amount of
remaining \HI\ that is spatially or kinematically distinct from the disk gas. 
In order to identify this faint emission we Hanning smoothed in 
velocity and spatially smoothed the data to twice the original beam size.  
The resulting cube has a
resolution of $28$\arcsec $\times22$\arcsec.  The rms noise improves to $\sigma=
0.15$ mJy beam$^{-1}$, or a 5$\sigma$ mass sensitivity of $1.7\times10^5$
\msun\ channel$^{-1}$ and column density of $8.9\times10^{18}$ 
cm$^{-2}$ beam$^{-1}$ channel$^{-1}$.  We visually searched this cube and found several 
examples of anomalous \HI\ features whose basic properties are presented in Table 
\ref{cloudprops}: their projected radial distance from the center of the galaxy, their 
velocity offset from the \HI\ disk, and their \HI\ mass as well as an estimate for their
contribution to the mass accretion rate.  This is not
an exhaustive list, but merely illustrates some of the most prominant examples
of HVC analogs associated with NGC~2997.  Figure \ref{mom0} shows where we have taken
slices through the smoothed data cube to present these features in Figures 
\ref{allslices}A--D.

Of these anomalous features, we find 5 examples of \HI\ clouds that are 
completely distinct from the \HI\ disk.  They were identified in p-v slices as 
having velocity offsets of greater than 50\kms\ from \HI\ gas in the disk at the same 
position (Figures~\ref{mom0} \& \ref{allslices}).  In three of the five examples, the 
\HI\ gas is counter-rotating with respect to the \HI\ disk (Slices A \& C), and in one 
instance we find an \HI\ cloud whose velocity component along the line of sight is
actually faster than the rotation of the disk (Slice B).  The clouds we identify
reside within 20 kpc from the center of the galaxy, which 
places them within the extent of the projected \HI\ disk, but outside the optical 
disk as it is seen in DSS images (Figure \ref{moment}), and they all have 
\HI\ masses of approximately $10^7$ \msun.  They were not detected by 
\citet{2007ApJ...662..959P} because of the low angular resolution of their Parkes (14\arcmin) 
and ATCA ($\sim1$\arcmin) data.  There are significant indications of a larger
population of \HI\ clouds around NGC~2997, however we are less confident in their
identification because they are closer to our mass sensitivity limit and because 
of residual phase calibration errors in the data cube.  

In addition to the \HI\ clouds, there are two other prominent features
of anomalous \HI\ we see in velocity channels:  a ``spur'' 
of counter-rotating gas near the center of the galaxy and a ``tail'' of emission lagging 
in rotation, extending to the west from the north half of the galaxy.  
Unlike the \HI\ clouds which are distinct, both of these features are connected to 
the main disk of the galaxy (making it difficult to estimate their mass).  The 
``spur'' is attached north
of the center, and arcs to the east and south extending greater than 6 kpc 
east from the center of the galaxy ($v=1016.1-1082.0$\kms\ 
in Figures \ref{chan1} \& \ref{chan2}). In p-v slices the same feature
looks like a shell of emission that is connected to the disk
(Figure \ref{allslices}B, C \& D). We estimate the \HI\ mass in the spur 
to be $\sim2.6\times10^7$ \msun.  The ``tail'' contributes to the asymmetry total
intensity map (Figure \ref{moment}).  Although the contours of the intensity 
weighted velocity map (Figure \ref{moment}) are well behaved, individual velocity 
channels, $v=943.5-983.1$\kms\ (Figure \ref{chan1}), show emission on the north 
side of the galaxy.  The coherence of this emission can be seen in Slices
A \& B of Figure \ref{allslices}: the feature looks like a bifurcation of the \HI\ 
disk in p-v space.  Its velocity component along the line of sight is actually moving faster 
than the rotation of the disk.  We calculate the mass of the tail to be $3.1\times10^7$
\msun.

\begin{table}[h!]
  \begin{center}
    \caption{Properties of Anomalous \HI\ Clouds in NGC~2997 \label{cloudprops}}
    \begin{tabular}{ccccccc}
      \tableline\tableline
      & Distance & Velocity\tablenotemark{a} & TF Mass\tablenotemark{b} & Accretion Rate\tablenotemark{c} & \\
      Obj. ID & (kpc) & (km s$^{-1}$) & (\msun) & (\msun) & (\msun\ yr$^{-1}$) & Notes \\
      \tableline
      Cloud 1 & 17.7  & 120  & $1.9\pm0.2\times10^7$  & 0.13 & Fig. \ref{allslices}A \\
      Cloud 2 & 13.0  & -105 & $7.6\pm0.1\times10^6$  & 0.06 & Fig. \ref{allslices}B \\
      Cloud 3 & 18.9  & 150  & $8.6\pm0.4\times10^6$  & 0.07 & Fig. \ref{allslices}C \\
      Cloud 4 & 14.2  & 80   & $8.9\pm0.8\times10^6$  & 0.05 & Fig. \ref{allslices}D \\
      Cloud 5 & 10.4  & 200  & $4.9\pm0.4\times10^6$  & 0.10 & Fig. \ref{allslices}A \\
      \HI\ Spur & 5.9 & 150  & $2.6\pm0.1\times10^7$  & 0.70 & Fig. \ref{allslices}B, C \& D \\
      \HI\ Tail & 27.8 & -60 & $3.1\pm0.3\times10^7$  & 0.07 & Fig. \ref{allslices}A \& B \\
     \tableline
    \end{tabular}
  \tablenotetext{a}{Negative indicates a line of sight velocity faster than the rotational
    velocity of the disk at that position.}
  \tablenotetext{b}{The accretion rate is estimated from \mhi$_{cloud} \times v_{offset}/r$ 
    where $v_{offset}$ is the velocity difference between the cloud and gas in the thin \HI\ 
    disk, and $r$ is the projected distance of the cloud from the center of the galaxy.}
  \end{center}
\end{table}

\section{Discussion}
\label{discussion}

In addition to the thin disk of NGC~2997, seen in most spiral galaxies, we 
have identified two additional \HI\ components that are not as ubiquitous:  a lagging,
thick disk and anamolous \HI\ clouds.  The thick disk has a mass of $1.4\times10^9$ \msun,
a broader velocity dispersion, $\sim16$\kms, and lags behind the rotational velocity
of the thin disk by $\sim60$\kms.  The anomalous \HI\ clouds are spatially or 
kinematically distinct from both \HI\ disk components.  The clouds have individual \HI\
masses between $10^{6.7-7.3}$\msun, are offset from the disk by $80-200$\kms\
and, in some instances, are counter-rotating.  In this section we compare our results
with the distribution of \HI\ observed around other galaxies, we speculate on 
the possible origins of NGC~2297's HVC analogs, we discuss our results in
the context of the Local Group HVCs, and we examine the effect of our 
distance estimate on our conclusions.

\subsection{Extraplanar \HI}
\label{extraplanarhi}

NGC~2997 is one of the lowest inclination galaxies in which a lagging disk has 
been modeled.  It is impossible to do this same decomposition for 
a face-on galaxy because the measured velocities contain no information on the 
rotation of the gas, but moderate to high inclination galaxies are well suited to 
this analysis.  The velocity gradient we estimate for NGC~2997, 
$18-31$\kms\ kpc$^{-1}$, appears to be typical of spiral galaxies for which 
deep observations reveal extended extraplanar gas.  Our findings are consistent 
with that seen in NGC~2403 where 
\citet{2001ApJ...562L..47F} calculate the thick disk to have a mean rotation velocity 
$25-50$\kms\ lower than that of the thin disk, in NGC~4459 where \citet{2005A&A...439..947B} 
find a thick disk that lags by $12-25$\kms, and in NGC~891, where the velocity 
gradient ranges from 14 to 43\kms\ kpc$^{-1}$ depending on the location in the galaxy
\citep{2007AJ....134.1019O}.  NGC~6946 also has a similar lagging halo of cold
gas \citep{2008arXiv0807.3339B}.

This range of values for the velocity gradient in spiral galaxies
suggests that one might correlate it with star formation activity.  
\citet{2006ApJ...636..181H,2006ApJ...647.1018H,2007ApJ...663..933H} have done 
a complementary study of three high inclination galaxies, NGC~5775, NGC~891, 
and NGC~4302, in which they find that the magnitude of the velocity gradient, 
measured in ionized gas (H$\alpha$), decreases with star formation 
activity, as measured in far infrared luminosity.  For NGC~2997, we calculate a 
star formation rate of $5\pm1$ \msun\ yr$^{-1}$ from 60 and 100 $\mu$m IRAS fluxes 
\citep[following][]{2002AJ....124.3135K,2003ApJ...586..794B} and $1.4$ GHz fluxes
\citep[following][]{2002AJ....124..675C,2003ApJ...586..794B}.  However, it is 
unclear how well NGC~2997 supports such a correlation: its values for the velocity gradient 
and scale height are close to the mean of those galaxies, while the star formation rate is
a bit high.  \citet{2007ApJ...663..933H} also suggest that the product of the velocity 
gradient and the scale height tends towards $20$\kms.  We estimate $dv/dh_z=12-21$\kms, 
which is consistent with their suggestion.

From a theoretical standpoint, the detection of a thick, lagging disk
is consistent with the idea developed by \citet{1980ApJ...236..577B} in which 
the Galactic HVCs are clouds condensing out of a hot Galactic corona fed by supernovae in the disk.  In his 
galactic fountain model, gas blown out of the disk by star formation
moves radially outward and slows down to conserve angular momentum as it moves 
vertically away from the mid-plane.  If the observed lag is only due to conservation of angular momentum then, in the
case of NGC~2997, gas that lags by 60\kms\ behind a disk rotating at 200\kms\ originated 
70\% closer to the center of the galaxy.  However, if hydrodynamic effects are important, the
radial motion is less dramatic.  Cooling gas condenses 
out of the corona and moves inward again.  Evidence for radial inflow of $10-20$\kms\ has been observed in NGC~2403
\citep{2001ApJ...562L..47F} and perhaps in NGC~4559 
\citep{2005A&A...439..947B}.  We do not find evidence for radial inflow of
\HI\ in NGC~2997, but such a signature may be ambiguous due to the low 
inclination of the galaxy.
Radially infalling clouds would also be expected if these clouds originated in a
cooling hot galactic halo that is at the virial temperature of the dark matter halo
as proposed by \citet{2004MNRAS.355..694M}, however
holes in the \HI\ disk may be evidence that supernovae are responsible for
circulating this gas.  

In a number of cases where we see anomalous \HI\ clouds in p-v slices, we also 
see depressions in the \HI\ column density in the disk (Slices B \& D).  These \HI\ ``holes'' 
occur near the outskirts of the \HI\ disk, but outside of the optical disk, and we
hesitate at calling them holes because they frequently occur between two \HI\ spiral arms.
These may be sites where \HI\ has been blown out of the disk from supernovae, 
however one cannot rule out that these may also be the result of a dwarf galaxy 
that has impacted the disk.  For example, this may be the case for large holes in NGC~6946 
\citep{2008arXiv0807.3339B} and M~101 \citep{1988AJ.....95.1354V}.  
In any case, it is interesting to note that all of the anomalous \HI\ we identify 
is coincident with the \HI\ disk and we do not find any large \HI\ clouds far
from it.

The large clouds of counter-rotating \HI\ gas and the ``spur'' cannot be explained by a galactic fountain
mechanism.  While fountain gas lags behind the main disk in rotation, it will conserve 
angular momentum and so can not be counter-rotating.  Additionally, to eject 
clouds of $10^7$ \msun\ from the disk at the observed velocity offsets requires an spatially 
and temporally coherent injection of $10^{54}$ ergs of energy. This corresponds 
to of order $10^4$ supernovae, assuming 10\% efficiency.  Despite the star formation rate of 
NGC~2997, this appears to be implausibly high.  Based on this energy argument, and their 
position with respect to the optical disk, we conclude that these \HI\ features are 
more likely material being accreted by NGC~2997, either out of a hot galactic halo 
\citep{2008ApJ...674..227P}, or are the remnants of a tidally disrupted gas rich dwarf galaxy.  
The maximum individual cloud mass from the \citet{2008ApJ...674..227P} simulations is 
$5\times10^6$ \msun, however they do not rule out the possibility of cloud complexes 
having higher masses.  We estimate a lower limit for the total
accretion rate of $1.2$ \msun\ yr$^{-1}$, which may help sustain the vigorous star formation
rate, especially if it is augmented by gas from the lagging disk which one would expect to 
eventually fall back towards the mid-plane.

Features such as the asymmetry in the extended northwest spiral arm are usually
interpreted as evidence for past interactions between galaxies, and there are two 
possible culprits.  The nearest LGG~180 group galaxy is a dwarf galaxy, ESO~434-G041,
which is $\sim115$ kpc to the southeast.  It is resolved in our ATCA observations, 
but lies outside the primary beam of the GMRT.  NED lists another dwarf galaxy, 
ESO~434-G030, $\sim90$ kpc to the southwest, but we do not see this galaxy in our data.  
Assuming that the velocity differences in the plane of the sky are similar to the 
line-of-sight velocity differences between these galaxies and NGC~2997, any encounter
would have taken place $\sim1$ Gyr ago.  While there are no apparent connections
between these galaxies and NGC~2997 in \HI, nor any optical signatures of disturbances
in the dwarf galaxies, a past tidal event or interaction seems to be the most likely
explanation for pulling such a large amount of \HI\ from the disk in such a coherent manner,
causing the asymmetry.  On the other hand, we do not believe that interactions are the most 
probable origin for the anamolous \HI\ clouds. 
Similarly, the ``tail'' feature, on the northwest side of the galaxy, seems best
explained by accretion--perhaps a dwarf galaxy that has been tidally disrupted as it passed near NGC~2997.

\subsection{Comparison with M~31 Clouds and Galactic HVCs}
\label{hvcnumbers}

We have identified the most massive anomalous \HI\ clouds associated with NGC~2297.
These five clouds have masses of order $10^7$ \msun.  The spectrum of cloud masses 
extends to lower masses but, in many cases, it is difficult to separate them as 
distinct from emission in the thick disk.  Nonetheless, we can still compare the clouds we 
have identified with those reported in the Local Group and around other galaxies.

The \HI\ mass distribution of clouds in M~31 given by dM(\mhi) $\propto$
\mhi$^{-2}$d\mhi, allows us to estimate the total number of similar
clouds we should expect to see around NGC~2997.  Most clouds have masses in 
the range $(0.15-1.3)\times10^6$ \msun\ with only Davies' Cloud
having \mhi\ $=10^7$ \msun.  For sources smaller than our beamsize, seen by 
\citet{2004ApJ...601L..39T} we would expect to detect about 10 of the M~31 HVC analogs 
at the 5$\sigma$ level (and about 4 above the 10$\sigma$ level); for those
clouds that we could resolve, we lack the column density sensitivity to 
detect them.  However, we already know from the observations of 
\citet{2007ApJ...662..959P} that there are no low column density \HI\ 
clouds residing off the disk of NGC~2997.  The HVCs discovered around M~31 in 
higher resolution images by \citet{2005A&A...436..101W} have an average size 
of 1 kpc (equivalent to the size of our beam at the distance of NGC~2997) and 
are below our detection limits.

With recently determined distant brackets to Milky Way HVCs, we can put our
\HI\ clouds into a local context.  For Complex C, \citet{2007ApJ...670L.113W}
 estimate \mhi\ to be $(0.7-6)\times10^6$ \msun.  The \HI\ masses 
for four other clouds range from $(0.09-0.41)\times10^4$ \msun\ 
(Cloud g1) to $(0.68-1.60)\times10^6$ \msun\ \citep[Complex GCP;][]{2008ApJ...672..298W}.
Therefore, at the distance of NGC~2997, we are only sensitive to the most 
massive Milky Way HVCs for which distance measurements exist.  However, we 
also see more massive features ($\sim10^7$ \msun) for which their are
few counterparts around the Milky Way and M~31.  Nonetheless, our 
discoveries are consistent with anomalous \HI\ clouds observed in other 
galaxies \citep[][and references therein]{2008A&ARv..15..189S}.

\subsection{Implications of the Distance Estimate to NGC~2997}
\label{distance}

The distribution of \HI\ clouds we find around NGC~2997 can be used to place
limits on cosmological and large scale structure simulations for the number of
small dark matter halos around massive galaxies.  The \HI\ masses we calculate
depend on the square of the assumed distance to the galaxy.  However, local large
scale structure results in an inaccurate distance estimate to NGC~2997 because
its peculiar velocity is not dominated by Hubble flow.  Unfortunately, an attempt to
correct for this, the velocity flow model of \citet{2005PhDT.........2M}, is sparsely 
populated by galaxies in the volume around NGC~2997, so we choose to calculate 
the distance to NGC~2997 using the Tully-Fisher relation (hereafter TF).

The TF relations based on near infrared observations may be the most 
reliable because they are less impacted by dust than optical magnitudes. 
We retrieved the 2MASS J, H, and K-band magnitudes from
NED and used the bias corrected TF parameters specific to Sc galaxies
\citep[][Table 2]{2008AJ....135.1738M}:
\begin{eqnarray}
\nonumber
M(J)-5\log h = -21.017 - 9.228(\log W-2.5)\\
\nonumber
M(H)-5\log h = -21.847 - 9.165(\log W-2.5)\\
\nonumber
M(K)-5\log h = -22.039 - 10.092(\log W-2.5)
\end{eqnarray}
To estimate the internal extinction, we interpolated between I and K bands to 
get values for J and H using the formulae in \citet{2007ApJ...671..203C}.  The 
FWHM of the H~I profile ($W=445$\kms) comes from Parkes 
observations \citep[][Pisano et al. 2009, in preparation]{2007ApJ...662..959P},
corrected for instrumental effects, turbulent motions, cosmological broadening, 
and inclination as per the formula in \citet{2008AJ....135.1738M}.  From these 
2MASS TF relations we derive a mean distance of $12.2\pm0.9$ Mpc.

This new distance estimate places NGC~2997 about 10\% closer to the Milky Way than
the estimates found on NED, corrected for peculiar velocity due to local density
enhancements, and 18\% closer than the velocity flow model of \citet{2005PhDT.........2M}.
This places somewhat tighter constraints on the fact that we do not see a large 
population of \HI\ clouds outside the disk of NGC~2997.  The anomalous \HI\ clouds within the disk,
for which we calculate masses, become more consistent with those observed around M~31 and more
consistent with the mass estimate for Complex C in the Milky Way.  Despite 
this closer distance estimate, the difference between it and previous estimates is not enough 
to change the formation hypotheses for the individual clouds we identify.

\section{Conclusions}
\label{conclusion} 

Our observations of NGC~2297 reveal a galaxy with a lagging, thick disk and anamolous 
\HI\ clouds in the halo similar to other galaxies, such as NGC~2403 
\citep{2001ApJ...562L..47F} and NGC~6946 \citep{2008arXiv0807.3339B}.  The presence 
of a lagging disk, which may be caused by star formation activity, and the 
presence of large \HI\ clouds and counter-rotating gas from accretion show that 
many processes contribute to the distribution of this cold gas. 

We propose that the lagging disk is the result of a galactic fountain process in which
gas is blown out of the disk by supernovae.  This gas experiences a decreasing gravitational 
force as it moves away from the mid-plane, so it drifts radially outward.  As a result, 
this gas rotates in the same sense as the thin disk, but at slower velocities to conserve 
angular momentum.  From IRAS 60 and 100 $\mu$m, and 1.4 
GHz radio continuum fluxes we estimate fairly vigorous star formation rate in NGC~2997 
of about 5 \msun\ yr$^{-1}$ to support the renewal of this lagging disk.  Of the total \HI\ 
mass of the galaxy, $16-17$\% resides in this lagging disk.

The most likely origin for the \HI\ clouds we discover is accretion from a hot galactic halo
or from nearby dark matter halos.  The \HI\ cloud masses are of order $10^7$ \msun.  From
their velocity offset with respect to nearby disk gas, we estimate a kinetic energy
for the clouds relative to the rotation of the disk of $\sim10^{54}$ ergs and conclude 
that it is unlikely their presence is due to supernovae.  Meanwhile, the spur and tail may be the
remnants of tidally disrupted gas rich dwarf galaxies.  The anomalous \HI\
we identify makes up approximately $1-2$\% of the total \HI\ mass.  Finally, the asymmetry of 
NGC~2997 to the northwest appears best explained by a tidal interaction
with another galaxy, although it is not obvious whom the culprit may be.

The features we find are not out of line with those observed in the Local Group, nor
compared to other galaxies for which similar observations and modeling have been done, although
the clouds of NGC~2997 tend to be more massive than the clouds for which we have distance
brackets in the Milky Way, even with an improved distance from the 2MASS Tully-Fisher relations
that place NGC~2997 $12-23$\% closer.  In spite of NGC~2997's relative isolation, we believe 
that a number of dynamic processes contribute to the sum of extraplanar \HI\ gas, 
including star formation, accretion, and galaxy interactions.  Deeper observations of 
NGC~2997 as have been accomplished with NGC~891 would likely continue to reveal extended 
structure.

\acknowledgements 
\section{Acknowledgements}
We gratefully acknowledge George Heald for sharing his modified GIPSY task to
model the lagging disk.  We thank the anonymous referee for his/her useful comments and constructive criticism that contributed to the improvement of this paper.  Finally, 
we thank the staff of the ATCA and GMRT who have made
these observations possible. The Australia Telescope Compact Array is part of
the Australia Telescope which is funded by the Commonwealth of Australia for
operation as a National Facility managed by CSIRO.  GMRT is run by the
National Centre for Radio Astrophysics of the Tata Institute of Fundamental
Research.  This work was supported by the National Science Foundation through 
grant AST~0708002.

\end{document}